\documentclass[letterpaper]{article} % DO NOT CHANGE THIS
\usepackage{aaai2026}  % DO NOT CHANGE THIS
\usepackage{times}  % DO NOT CHANGE THIS
\usepackage{helvet}  % DO NOT CHANGE THIS
\usepackage{courier}  % DO NOT CHANGE THIS
\usepackage[hyphens]{url}  % DO NOT CHANGE THIS
\usepackage{graphicx} % DO NOT CHANGE THIS
\usepackage{acronym}
\usepackage{natbib}  % DO NOT CHANGE THIS AND DO NOT ADD ANY OPTIONS TO IT
\usepackage{caption} % DO NOT CHANGE THIS AND DO NOT ADD ANY OPTIONS TO IT
\usepackage{algorithm}
\usepackage{algorithmic}

\usepackage{newfloat}
\usepackage{listings}
\DeclareCaptionStyle{ruled}{labelfont=normalfont,labelsep=colon,strut=off} % DO NOT CHANGE THIS
\floatstyle{ruled}
\newfloat{listing}{tb}{lst}{}
\floatname{listing}{Listing}
\title{A Chain Is Only as Strong as Its Weakest Link: A Scoping Review of System Integration Audits in AI}
\author{
    Leah Davis\textsuperscript{\rm 1},
    Dominic Martin\textsuperscript{\rm 2},
    AJung Moon\textsuperscript{\rm 1}
}
\affiliations{
    \textsuperscript{\rm 1}McGill University, Montréal, QC\\
    \textsuperscript{\rm 2}Université du Québec à Montréal, Montréal, QC\\
    leah.davis@mail.mcgill.ca, 
    martin.dominic@uqam.ca,
    ajung.moon@mcgill.ca
}

\usepackage{bibentry}
\acrodef{AI}{artificial intelligence}
\acrodef{ML}{machine learning}
\acrodef{GPAI}{general-purpose artificial intelligence}
\acrodef{EU}{European Union}
\acrodef{LLM}{large-language model}
\acrodef{LLMs}{large-language models}
\acrodef{APIs}{application programming interfaces}
\acrodef{FRT}{facial recognition technology}
\acrodef{SAE}{Society of Automotive Engineers}
\acrodef{FAA}{Federal Aviation Administration}
\acrodef{ISO}{International Organization for Standardization}
\acrodef{IEEE}{Institute of Electrical and Electronics Engineers}
\acrodef{SAE}{Society of Au-
tomotive Engineers}

\begin{document}

\maketitle

\begin{abstract}
%As artificial intelligence (AI) systems have expanded into complex configurations of interdependent components. Yet the component-level audits dominating the evaluation landscape are inadequate to protect against risks arising from the integration and interactions between components, external deployment environments, and multi-AI system architectures. Since the 1960s, \textit{system integration} has been central to software evaluation practice, serving as a foundational element of audits in safety-critical domains. Comparable lines of work have emerged in the AI audit literature, but the degree, scope, and consistency with which these audits are designed or applied within AI remain unknown.% 

As AI systems become increasingly integrated into diverse interfaces and applications, model-centric audits are insufficient to address risks arising from interactions among system components and deployment environments. System integration has long been central to software audits in safety-critical domains such as aerospace. However, its role in AI auditing remains underexplored. Scanning through 4,259 documents, we present a scoping review of AI audits that treat \textit{system integration} as a core tenet of evaluation (n = 58). Using reflexive thematic analysis, we analyze their elements, actors, enablers, and constraints. We find that the corpus represents an emerging yet still fragmented form of AI auditing: few existing measures target integration-specific risks; large gaps remain in meeting traditional audit expectations; and access to necessary information and resources significantly influences audit design. Nonetheless, integration can be categorized across three sites (inter-component, system-environment, and multi-system), each serving the functions of risk exploration, risk determination, coordination, and procedural regularity. Deviating from other types of evaluations, these audits assess qualities specific to system integration, including compatibility, completeness, and oversight. This review calls on the AI community to prioritize system integration as a core strategy for addressing AI risk, and to develop audit practices capable of capturing failures across components, environments, and systems beyond the reach of component-level evaluation.

%This review contributes: (1) an account of the coverage system integration audits in AI provide; (2) their relative placement within the broader AI evaluation ecosystem; and (3) factors relevant to institutionalizing this distinct audit class. 

\end{abstract}


\section{Introduction}

In 2018 and 2019, two Boeing 737 MAX 8 aircraft crashed shortly after take-off, resulting in 346 fatalities \cite{malunga_key}. Two \textit{system integration failures}--defined as breakdowns in the coordinated functioning of interconnected components, subsystems, and the actors responsible for them--were primary contributors to the tragedy  \cite{madni_systems_2014}. Incident analyses report (1) the technical integration of the aircraft's control software was updated and deployed without adequate review \cite{nartey_boeing_2025, nicas_boeing_2019, prokop_software_2024}; and (2) the \textit{sociotechnical nature} of the aircraft as a system was undermined, resulting in the pilots’ lack of training in countering related errors \cite{appicharla_boeing_2023, lengyel_examining_2023}. Notably, incidents like these are rare in highly regulated industries such as aerospace that understand safety as part of a \textit{system}; accordingly, component and subsystem integration is subject to dedicated audits that we collectively define as \textit{system integration audits} \cite{pop_enhancing_2023}. 
% The aerospace industry has highly regulated audit regimes designed specifically to prevent failures in system integration, but the Boeing incidents illustrate what can occur when these safeguards are not properly applied or enforced \cite{pop_enhancing_2023}.
%

The presence of these incidents in such an established industry foreshadows a stark warning for the \ac{AI} community, an integrated ecosystem where comparable safeguards have yet to fully develop against ad-hoc quality expectations. %Today, the evaluation of \ac{AI} systems against quality expectations remains ad hoc. 
% Artificial intelligence (AI) represents one such case: a field showing overwhelming evidence of harm when experimentation and profitability push testing priorities aside. 
% Integration failures involving AI will only be amplified as the popularity of general-purpose artificial intelligence (GPAI) systems --reusable foundation models that can be embedded across diverse downstream systems and countless applications --grows. 
% AI integration increasingly shapes vast socioeconomic systems, yet many of these integrations occur with limited transparency into how the underlying components, sub-systems, and surrounding environments function. 
This is especially problematic in the context of \ac{GPAI}, where a small number of oligopolistic upstream providers supply powerful foundation models as a single component integrated into millions of downstream systems \cite{williams_regulating_2025}. In many cases, system integrators may even bear responsibility for system failures, despite having little control over the upstream technologies on which they depend. For example, the point at which an integrator’s modifications become significant enough to trigger legal obligations remains ambiguous under the \ac{EU}'s AI Act~\cite{ebers_upstream_2026}. Related vulnerabilities also extend beyond legal responsibility to the \textit{management} of system-level faults, where risks extend across the broader ecosystem, affecting end-users, and the auditors responsible for evaluating these systems. \textit{Belo}, a financial technology start-up based in Argentina, provides a cautionary tale of such risks. Without warning, the firm discovered that its access to Anthropic's Claude had been revoked, disrupting service integration for more than 60 of its downstream accounts--with nothing but a Google Form for recourse \cite{anthropics}. The case illustrates a broader problem: the abundance of model-centric evaluation practices is not suited to discover and prevent failures 
% Yet the explicit system integration audit protocols capable of targeting these vulnerabilities do not exist in AI. Evaluation practices remain component-specific and model-centric, providing limited coverage 
across sites where system interactions occur.

To examine these gaps, we conduct a scoping review exploring the conceptualization and application of system integration audits in AI across a broad set of literature. As a first step in consolidating this work and building a foundation for system-level AI assurance, we address three questions:
%Moreover, the term “audit” amongst AI evaluation departs significantly from the standardized meaning in other fields, and the diversity of approaches labelled as such raises questions about the level of assurance they can meaningfully provide. Though the existing landscape has yet to be systematically examined, system integration audits present a promising avenue for strengthening AI governance.%

\begin{itemize}
  \item \textbf{RQ1} Where in AI systems do common system types, application areas, and integration sites emerge as potential targets for integration-specific audits?
  \item \textbf{RQ2} How does the auditing of these sites provide, or fail to provide, coverage across AI systems?
  \begin{itemize}
      \item \textbf{RQ 2.1} Which roles are involved?
      \item \textbf{RQ 2.2} What qualities do they evaluate?
      \item \textbf{RQ 2.3} What are the audit outputs and related functions they produce?
  \end{itemize}
  \item \textbf{RQ3} Which factors must be considered in institutionalizing system integration audits in AI?
\end{itemize}

% This review contributes a comprehensive mapping of 58 research documents. 

In this article, we first outline what system integration audit practices are in established, safety-critical industries and relate them to integration-related vulnerabilities and regulatory gaps in \ac{AI} (Section 2). Following the method of the scoping review (Section 3), we articulate the results (Section 4) across four areas of system integration audits: the three central sites where they are conducted, how audit roles across these sites deviate from traditional expectations of independent audits, the qualities assessed, and the four functions they serve. Section 5 synthesizes these findings, emphasizing that despite an emerging recognition of AI risks arising through system integration, system-level audit expectations remain underdeveloped and constrained by limited access to information, resources, and technical evidence. Section 6 concludes by outlining steps to advance these audits in practice.

\section{Background}
\subsection{Existing System Integration Audits}

System integration audits are applied across many \textit{systems}--what seminal scholar Donella Meadows defines as “an interconnected set of elements that...must consist of three things: elements, interconnections, and a function or purpose” (\citeyear{meadows_thinking_2008}, p. 12). In the medical device industry, supplier audits (\ac{ISO} 13485) ensure the integration of components from external vendors is satisfactory \cite{noauthor_international_2025, smogavc_cestar_mastering_2023}, with military software systems similarly relying on configuration and subcontractor integration audits \cite{summers_effective_2013}. Within the aerospace industry, the audit class has been established since the 1960s \cite{zieke_progress_1963}. The primary function of these audits is to concretize the discovery, management, and prevention of the unintended behaviours and coupled dependencies arising from component and system interactions. This is especially important for highly salient points of software and hardware integration among environmental control systems (cabin pressure), electronic and hydraulic suites (landing gear), and mechanical elements (fuel injection) \cite{moir_aircraft_2011}. 

Through industry standards such as the \ac{SAE} Aerospace Recommended Practice 4754B and oversight from accreditation bodies including the International Civil Aviation Organization, \ac{EU} Aviation Safety Agency, and the \ac{FAA}, these audits have become well-institutionalized protocols mandatory for all aircraft \cite{automotive_engineers_arp4754a_2026, pop_enhancing_2023}. Table~\ref{tab:integration-faults} lists several common system integration faults that these audits evaluate, summarized in a case study by Winter (\citeyear{winter_system_2020}) on aerospace system development, alongside our interpretation of their associated risks. 
\begin{table}[t]
\centering
\setlength{\tabcolsep}{3pt}

{\small
\begin{tabular}{|p{\dimexpr0.35\columnwidth-2\tabcolsep\relax}|p{\dimexpr0.65\columnwidth-2\tabcolsep-3\arrayrulewidth\relax}|}
\hline
\textbf{Integration Fault}& \textbf{Associated Risks} \\
\hline
\textbf{Poor sequencing} 
(e.g., transitions between components and systems)&
Propagation of software failures, loss of system control, operational shutdowns in critical infrastructure \\
\hline
\textbf{Inadequate fail-safe margins} 
(e.g., the addition or removal of components or systems reduces safety margins)&
Increased likelihood of accidents, reduced tolerance for software or sensor errors, greater vulnerability to environmental disturbances \\
\hline
\textbf{Improper indication signalling} 
(e.g., relevant actors and parties are not alerted to system faults)&
Delayed responses to critical system failures, decisions made by auditors or end-users based on incomplete information, escalation of manageable failures into major incidents \\
\hline
\textbf{Unanticipated outputs from unknown inputs} 
(e.g., unforeseen stakeholder responses or emergent system functions)&
Unsafe automated responses, unclear attribution of liability, erosion of public trust \\
\hline
\end{tabular}
}
\caption{Common system integration faults detailed by Winter (2020), followed by our analysis of associated risks.}
\label{tab:integration-faults}
\end{table}
To address these risks, the aforementioned certification processes apply several audit conditions: (1) independent, external auditors; (2) systematic and traceable processes in accordance with a standard or certification framework; and (3) evidence tested at the same level as the requirements and faults themselves (system-level faults necessitate system-level tests) \cite{pop_enhancing_2023,winter_system_2020}. Together, these conditions promote rigour, enabling credibility, legitimacy, and due process. By contrast, AI systems, which remain vulnerable to comparable integration-related faults, remain largely unprotected by the same set of audit conditions.

\subsection{Integration-Related Vulnerabilities in Existing AI Evaluation}

Contemporary AI systems are rarely stand-alone: they are typically embedded in, or orchestrating the behaviour of, complex digital infrastructures \cite{varoquaux_hype_2025}. In earlier software development, when algorithmic systems operated on in-house machines or single-vendor stacks, assurance could be more easily provided for task-specific products or services. Today, \textit{independently} developed software and hardware components interact and share significant dependencies. In AI systems, this reliance is increasingly organized around oligopolistic GPAI models--a dominant form of system integration.

The EU AI Act defines GPAI \textit{models} to be, “...trained with a large amount of data using self-supervision at scale, that displays significant generality and is capable of competently performing a wide range of distinct tasks regardless of the way the model is placed on the market, and that can be integrated into a variety of downstream systems or applications.” (Article 3, Paragraph 66, Regulation (\ac{EU}) \citeyear{regulation_european_2024}/1689) Upstream providers supply \ac{LLMs}, text-to-image models, and those tailored for cyber-physical purposes \cite{brohan_rt-2_2024}. Once a GPAI model is embedded downstream into a deployment environment by a system integrator, a GPAI \textit{system} is created (Article 3, Paragraph 63, Regulation (\ac{EU}) \citeyear{regulation_european_2024}/1689). From decision-support tools in medical chatbots to software stacks in power grid management, these systems introduce new pathways for external impacts \cite{chow_large_2025, xie_massively_2022}. As a result, compliance and regulatory regimes are increasingly concerned with faults stemming from a \textit{system’s} overall behaviour, rather than the model itself \cite{carey_regulating_2025}. As model-as-a-service becomes more popular, GPAI integration has become relatively straightforward and inexpensive, drastically expanding the number and diversity of integrators and applications \cite{gan_model-as--service_2023}. In principle, such democratization of access to powerful capabilities is welcome. In practice, however, as Williams et al. (\citeyear{williams_regulating_2025}) note in their work on downstream regulation, more parties are exposed to liability risks, further fragmenting evaluation and accountability practices across the AI industry.

A significant contributing factor is the industry’s over-reliance on component-level evaluations, which restrict the assessment scope to a single \ac{AI} pipeline element \cite{varoquaux_hype_2025}. Rismani et al. (\citeyear{rismani_measuring_2025}) report that over 57.3\% of responsible AI measures target the model, while only 3\% do so for the full system. Similarly, Carey (\citeyear{carey_regulating_2025}) observes that a major regulatory blind spot in the \ac{EU} AI Act is its continued emphasis on risks stemming from \ac{GPAI} \textit{models} rather than \ac{GPAI} \textit{systems}. As a result, the information needed to prevent harms arising from integration faults (see Table~\ref{tab:integration-faults}) remains largely neglected. 

Surprisingly, open-source, upstream \ac{GPAI} model providers are subject to surprisingly relaxed regulatory expectations. The \ac{EU} AI Act, for instance, spells out only two obligations that directly apply to \ac{GPAI} systems: outputs must be marked as artificially generated in a machine-readable format and providers must cooperate with enforcement authorities \cite{szadeczky_risk_2025}. Neither obligation addresses the disproportionate burden of compliance duties on downstream actors. Without \ac{AI} governance mandates specifying how responsibilities, information, and access concerns should travel down the supply chain, downstream actors are left with suboptimal means to examine the performance and safety issues they inherit. While model-centric evaluations remain useful for benchmarking unit tests, they provide little insight into risks arising from interactions and dependencies between components and the environment. As Correll et al. (\citeyear{correll_navigating_2014}) note in their review of evaluation approaches, “...as reductionist techniques attempt to control for more and more components, they become less and less adept at validating particular real-world systems.” (p. 2) 

\subsection{Evaluation Gaps Surrounding AI Auditing}

Regulators have begun to recognize these system-level dynamics. For instance, Chapter V of the \ac{EU} AI Act is dedicated to establishing evaluation expectations among model providers and system integrators, calling for audits to ensure quality management (Annex VII, Regulation \ac{EU} \citeyear{regulation_european_2024}/1689; \citealp{bieger_evaluation_2016, rico_general-purpose_2025}). Yet the regulation itself offers little guidance on how such audits should be operationalized: where they should occur within an AI system, who should be involved, and the qualities or functions they should evaluate. Unlike the aerospace industry's highly structured audit regimes, AI auditing has emerged without clear mandates, leading to diverse approaches that vary widely in structure, scope, and methodological rigour \cite{birhane_ai_2024}. As Vecchione et al. (\citeyear{vecchione_algorithmic_2021}) note, “...the meaning of the term has become ambiguous, making it hard to pin down what audits actually entail and what they aim to deliver.” (p. 1) Yet, the audit conditions traditionally applied, such as independent oversight, are often absent in practices \textit{labelled} as AI audits \cite{birhane_ai_2024}. If governance frameworks expect audits to deliver holistic assurance across AI systems, greater clarity is needed in how they are, and are not, doing so.

\section{Methodology}
To provide a landscape overview of the ways in which system integration is operationalized in AI, we conducted a scoping review using the Preferred Reporting Items for Systematic reviews and Meta-Analyses extension for Scoping Reviews (PRISMA-ScR) \cite{tricco_prisma_2018}. The main inclusion and exclusion criteria are as follows:

\begin{enumerate}
  \item \textbf{Reference to audits:} The document must propose, conduct, assist with, improve upon, or review an audit or audit artifact. Evaluations, such as benchmarks, were excluded \textit{unless} they mentioned auditing. In this review, we avoid prescribing what an AI audit should mean or discriminating among different auditing approaches. 
  \item \textbf{Relevance to AI/\ac{ML}:} The object or target of the audit must be an AI system. Documents using AI solely as a tool to evaluate another system were excluded.
  \item \textbf{Relevance to system integration:} The audit must examine the integration, interaction, or interconnections among AI \textit{systems} beyond the model. 
  \item \textbf{Publication type:} We included published articles; e-book or book chapters; industry white papers; and pre-prints. \ac{ISO} standards and those from the \ac{IEEE} were excluded due to inaccessibility or underdevelopment.  
  \item \textbf{Language:} We included only research documents written in full English.
\end{enumerate}

We did not discriminate against literature based on publication date or geographical location. 

\begin{figure*}[t]
    \centering
    \includegraphics[width= 0.70\textwidth]{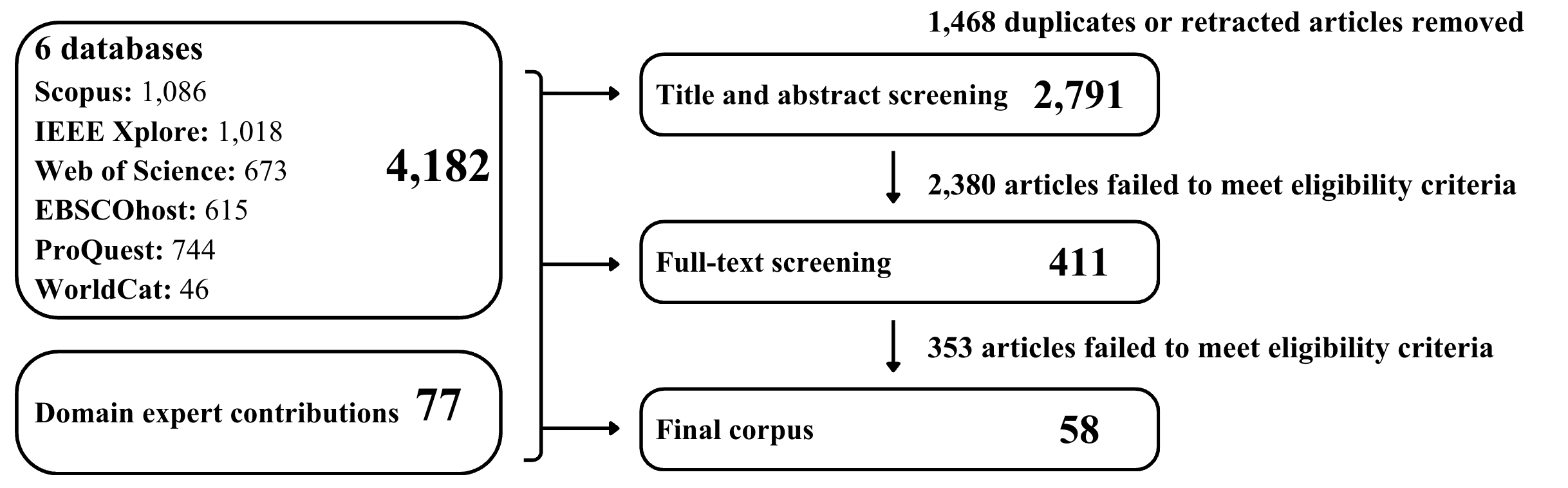}
    \caption{The Preferred Reporting Items for Systematic reviews and Meta-Analyses extension for Scoping Reviews flow chart of document inclusion.}
    \label{fig:PRISMA}
\end{figure*}

\subsection{Database Search Strategies and Study Selection}

To capture a broad perspective of system integration across disciplines, we queried six computing and social science databases: \textit{Scopus}, \textit{\ac{IEEE} Xplore}, \textit{ProQuest}, \textit{Web of Science}, \textit{EBSCOhost}, and \textit{WorldCat}. Searches took place between mid- and late November 2024. Limiting our search to the abstract, title, and keyword fields, our queries sought at least one of “algorithmic”, “artificial intelligence”, “ai”, “machine learning”, “ml”, and at least one of “audit”, “audits”, “auditing”, “auditor”, “auditors” occurring within a three-word proximity. It was not possible to directly search for “system integration auditing" because the term is not used systematically, returning too many unrelated results. Rather, we analyzed a broad set of literature claiming an auditing approach, then narrowed the corpus. The initial queries resulted in 4,182 documents (Scopus: 1,086, \ac{IEEE} Xplore: 1,018, ProQuest: 744, Web of Science: 673, EBSCOhost: 615, WorldCat: 46). An additional 77 documents were manually added from the authors’ existing collection of \ac{AI} auditing literature, totalling 4,259 documents for screening. Figure \ref{fig:PRISMA} presents the PRISMA-ScR flow diagram. Details of the query results are included in Appendix A (Table A1).
 
The screening was performed using the Covidence platform. First, duplicates (1,443) and retracted (25) documents were automatically removed. An extensive screening phase then took place between December 2024 and February 2025. The lead reviewer screened the titles and abstracts of all documents against the eligibility criteria (Section 3.1) with the help of two other reviewers, who reviewed the same. Two reviewers voting “yes” or “no” on a document determined whether it was included in the full-text review. Documents with a “maybe” or a mismatch of votes were discussed by all three reviewers toward a final decision, with an inter-coder agreement of 82\% for the first 50 documents. The reviewers met every 100 documents to manage conflicts. 

The final corpus comprised 58 documents for qualitative coding, imported into \textit{ATLAS.ti} for analysis. Included studies are listed in Appendix B (Tables B1 to B4). One-third of the corpus was randomly sampled and independently coded by all reviewers to form an initial coding tree. Reviewers met every 3 to 5 documents to reorganize, merge, edit, and remove codes and broader code groups. Several code groups were mandated for the review, including definitions relevant to “audit” and “system integration”; the system type and application area; the \ac{AI} system elements implicated; the roles involved; the qualities assessed; and the audit outputs and functions. The lead reviewer then completed a second round of qualitative coding across the corpus, re-coding the initial one-third of documents. 

\section{Results}
The corpus suggests growing attention to system integration as a core tenet of AI auditing, but across a fragmented body of work with ambiguous terminology, audit structures, and methodological commitments. The 58 documents are concentrated in peer-reviewed articles (43.1\% journal, 37.9\% conference), with the remainder comprising industry white papers (5.2\%), pre-prints (5.2\%), book chapters and workshop papers (3.4\%), and a policy report. The geographic distribution of the corpus also points to breadth. Where geographic information was reported, audit data collection and conduct took place in many countries and regions, including China, India, New Zealand, Cyprus, the United States, and Western European countries. This variation highlights that integration-related concerns are appearing across multiple jurisdictions and institutional settings, unconfined to a single regulatory jurisdiction.

The publication timeline further suggests increasing momentum, with the earliest publication dating to 2019 and a steady annual increase through 2024, which accounted for the largest proportion of the corpus at 29.3\%. This growth suggests that system-level evaluation is increasingly considered as AI systems are customized and integrated into a variety of operational environments. Yet the increase in publication volume does not indicate methodological maturity. Nearly two-thirds of the documents (65.5\%) conceptualize or propose an audit approach rather than apply an existing method. These proposals span more than 28 unique approaches, including auditing frameworks, protocols, mobile applications, \ac{APIs}, schemas, ontologies, simulations, case studies, catalogues, scorecards, and plug-ins. While the majority of these approaches target purely algorithmic systems (82.8\%), there is diversity in the interfaces audited, including embodied devices (e.g., physical entities without degrees of freedom, such as Internet of Things devices) at 10.3\%, robotic systems (e.g., possessing degrees of freedom in their movement) at 6.9\%, and simulated AI at 5.2\%. 

Fragmentation is also visible across the AI system types and application areas audited, suggesting that system integration concerns extend beyond any single technical architecture or context. The documents discuss 10 algorithm types, including recommendation algorithms, image classifiers, and knowledge graph networks. One-tenth of the documents state they can be applied to any algorithm. Despite mounting calls for broader system-level \ac{AI} audits, \ac{GPAI}, generative, and multi-modal systems together account for only 29.3\% of the corpus. Of the ten application areas identified, healthcare applications (e.g., clinical algorithms, medical AI platforms, and surgical robots) are most dominant at 34.5\%, likely due to their safety-critical nature, risk aversion, and low tolerance for failure. Hiring follows at 12.1\%, with transportation, finance, news/social media, and public service following at 5.2\% each, and \ac{FRT}, technology services, and advertising at 3.4\% each. 31.0\% of the documents discuss application-agnostic audits. Note that each system type, or application area, was coded more than once if the audit explicitly targets multiple within the document.

\subsection{Where Audits Take Place: Integration Sites Among AI Systems}
To examine how auditing practices could systematically extend beyond the model, we explore how \textit{system boundaries}--determining the elements included in the system of interest--are drawn. We identify six elements involved in integration: data/input, model, output, user-interaction, deployment, and ecosystem. The first four are common to the \ac{AI} development pipeline, and the remaining two lie external to it. See Appendix C for a description of each element.  

\begin{figure*}[t]
    \centering
    \includegraphics[width= 0.67\textwidth]{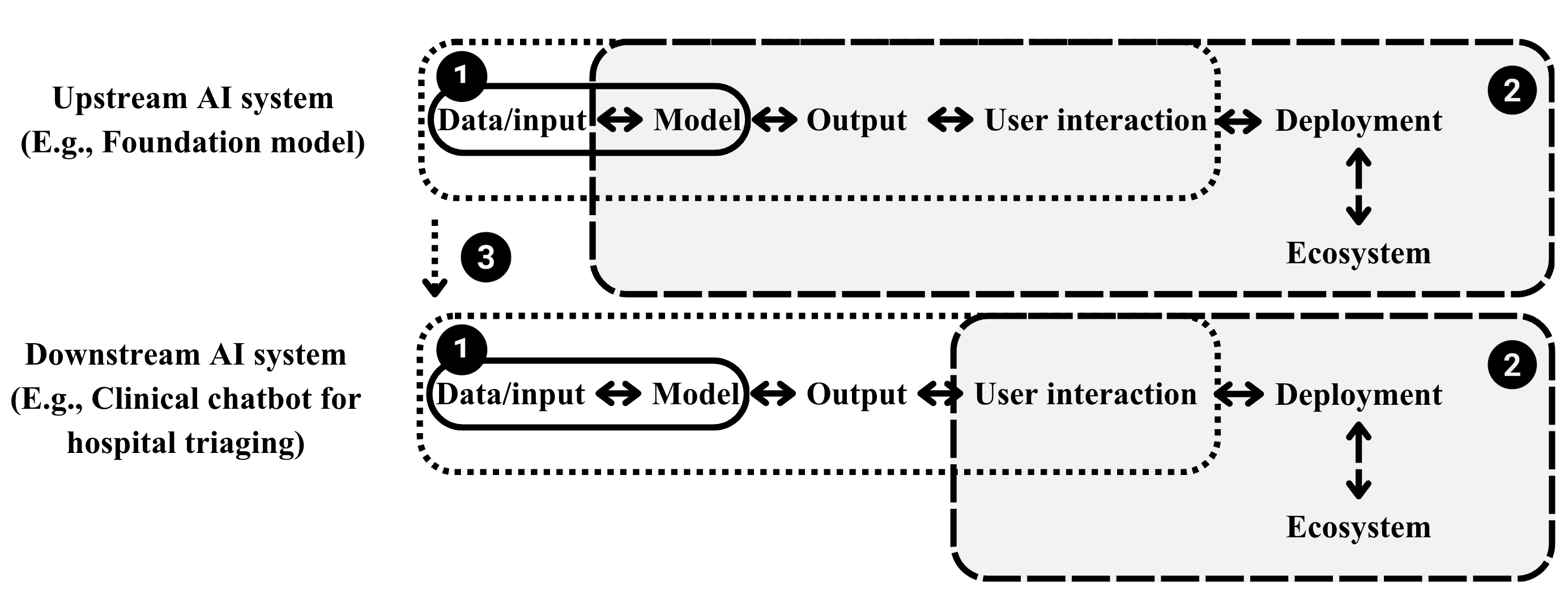}
    \caption{Three system integration sites are identified in this review: (1) inter-component integration (solid lines), (2) system-environment integration (dashed lines and shaded areas), and (3) multi-system integration (dotted lines).}
    \label{fig:integration_sites}
\end{figure*}

Following the definition of system integration as the coordinated functioning of interconnected components, subsystems, or systems \cite{madni_systems_2014}, we identify three primary integration sites: the points where these elements are integrated with one another. These include: (1) \textit{inter-component integration}, occurring between two or more pipeline components, (2) \textit{system-environment integration}, occurring between a single AI system and an external environment(s), and (3) \textit{multi-system integration}, occurring between two or more AI systems. Inter-component integration is the most granular in scope, targeting the four development components, whereas system-environment integration sites target the deployment and ecosystem elements. Multi-system integration has the broadest scope, covering all elements. Figure \ref{fig:integration_sites} provides a high-level overview of the sites and their relative positions among upstream providers and downstream integrators. In the following section, each site, its system boundary, and the \textit{intermediaries} engaged, defined by Madni and Sievers (\citeyear{madni_system_2014}) as the mechanisms (e.g., processes, tests) enabling the flow of information and resources across these boundaries, will be analyzed.\\

\textbf{(1) Inter-component integration:} \textit{E.g., An audit aims to examine whether API transmission of patient electronic health records to a diagnostic model introduces data formatting errors that affect model outputs.}\\

As detailed in the conceptual example, the system boundary encapsulates two or more development components (e.g., data/input, model) within a \textit{single AI system}, whether upstream or downstream. In the corpus, 87.9\% of documents engage the data/input component, followed by the model (67.2\%), the output (65.5\%), and user-interaction (48.3\%). Integration occurs in a linear, sequential fashion (i.e., data-to-model, model-to-output), indicating the importance of order. As a result, more traceable input-output (i.e., cascading) relationships are found. The majority of intermediaries are technical, akin to what traditional system integration audits in aerospace assess (Table~\ref{tab:intermediaries-integration-sites}). This site aligns with Rajabalinejad's (\citeyear {rajabalinejad_safe_2019}) definition of subsystem integration.

\begin{table}[h!]
\centering
\setlength{\tabcolsep}{3pt}

{\small
\begin{tabular}{|p{\dimexpr0.35\columnwidth-2\tabcolsep\relax}|p{\dimexpr0.65\columnwidth-2\tabcolsep-3\arrayrulewidth\relax}|}
\hline
\textbf{Integrated Components} & \textbf{Intermediary Examples} \\
\hline

\textbf{Data $\rightarrow$ Model} &
Datasheets, data-sharing agreements, human-annotation platforms, data acquisition systems, schema-enforcing systems, and legacy systems \\
\hline

\textbf{Model $\rightarrow$ Output} &
Model cards, application programming interfaces, and microservices \\
\hline

\textbf{Output $\rightarrow$ User-interaction} &
Emulators, simulators, wrappers, system prompts, and graphical user interfaces \\
\hline
\end{tabular}
}

\caption{Intermediaries in inter-component integration sites.}
\label{tab:intermediaries-integration-sites}
\end{table}

\textbf{(2) System-environment integration:} \textit{E.g., An audit aims to evaluate how physicians incorporate a hospital's AI diagnostic support system's recommendations into patient triage workflows.}\\

These audits evaluate a single AI system’s integration with an external environment(s). These audits are most common in downstream task-specific systems, but as Figure~\ref{fig:integration_sites} shows, they can also apply to upstream AI systems if used directly without modification. Contrary to inter-component sites, which reduce an AI system into discrete components, these audits take a constructivist stance, treating an AI system as a whole, “...since a system’s characteristics may be different than an amalgamation of the characteristics of its parts” (Gursoy and Kakadiaris \citeyear{gursoy_system_2022}, p. 9). Instead, the system boundary encompasses a deployment environment (found in 51.7\% of documents) and/or broader sociotechnical ecosystem (77.6\%).

Due to their external-facing nature, fewer input-output relationships are found. The integration is not sequential; instead, it is predominated by coupling effects and feedback loops involving stakeholders, resources, and occasionally, physical space. A greater diversity of audit structures thereby emerges. Intermediaries exhibit a much stronger sociotechnical concentration, functioning as mechanisms that enable the assessment of “the final, social process in which the decision making is embedded” (Hauer et al. \citeyear{hauer_assuring_2021}, p. 2). Intermediaries include sandboxes, post-market surveillance analyses, scenario testing, feasibility and ethnographic studies, end-user questionnaires or surveys, and conformity, risk, and impact assessments. A stronger process orientation is adopted, emphasizing documentation, feedback, and available remedy mechanisms. These sites could be described as forms of “human-systems integration” (Madni and Sievers (\citeyear{madni_systems_2014}), p. 48).\\
 
\textbf{(3) Multi-system integration:} \textit{E.g., An audit aims to assess how patient-record summaries generated by a commercial LLM (a system) feed into a clinical risk prediction model (another system) used for treatment decisions.}\\

These audits describe the integration of two or more AI systems, emphasizing a clear upstream-to-downstream directionality. In the corpus, these sites occur when a foundation model is integrated into an application-specific AI system. Unlike the first two sites, which are not concentrated on any particular system type, these sites typically involve \ac{GPAI}. With the system boundary encompassing both inter-component and system-environment sites, internal development components and external deployment and ecosystem stages are implicated, as well as the aforementioned technical and sociotechnical intermediaries (Table~\ref{tab:intermediaries-integration-sites}). At the broadest scope, these sites concern the design of AI infrastructure, including power, control, and access dynamics. Sub-themes of information and material flows, in relation to supply chains and their actors, are dominant. In safety engineering literature, this is referred to as system-of-systems integration \cite{madni_system_2014}. 

All three integration sites are not mutually exclusive; they are complementary. This simultaneity is illustrated by Mökander et al. (\citeyear{mokander_blueprint_2025}), who audit generative AI systems using three sub-audits. In the \textit{model sub-audit}, the data's impact on the model’s design and training is considered (inter-component integration). In the \textit{application sub-audit}, the impact of the downstream, task-specific AI system is assessed (system-environment integration), and in the \textit{governance sub-audit}, model access and dissemination strategies are reviewed (multi-system integration). To an extent, these sites provide natural leverage points in the system where intervention (e.g., further testing or remedy) can occur.

\begin{table*}[t]
\centering
\setlength{\tabcolsep}{3pt}

{\small
\begin{tabular}{|p{0.16\textwidth}|p{0.15\textwidth}|p{0.19\textwidth}|p{0.14\textwidth}|p{0.15\textwidth}|p{0.15\textwidth}|}
\hline
\textbf{Integration Site} &
\textbf{Location} &
\textbf{System Boundary Encapsulates} &
\textbf{Form of Intermediaries} &
\textbf{Commonly Associated Roles} &
\textbf{System Type Targeted} \\
\hline

\textbf{Inter-component} &
Internal development &
Data/input, model, output, and user-interface components &
Primarily technical &
Developers &
Any system \\
\hline

\textbf{System-environment} &
External user-facing environments &
Deployment and broader ecosystem elements &
Primarily sociotechnical &
Domain experts and end-users &
Any system \\
\hline

\textbf{Multi-system} &
Both internal and external settings &
All elements &
Both &
Management and policymakers &
Predominantly general-purpose, generative, and physical AI systems \\
\hline
\end{tabular}
}

\caption{Summary of integration site characteristics in artificial intelligence systems.}
\label{tab:summary}
\end{table*}

\subsection{Who is Involved in the Audits: Role Types}

After mapping where these audits take place, we explored who was involved--those with a significant contributory role in the audit. The corpus refers to a diverse set of at least ten distinct roles: auditors (44.8\%), developers (46.6\%), domain experts (44.8\%), end-users (39.7\%), management (10.3\%), policymakers (17.2\%), external vendors (25.9\%), system operators (13.8\%), automated AI and robotic agents (5.2\%), and a miscellaneous group (6.9\%). In general, the roles are ambiguous, lacking clarity about how individuals or organizations are expected to fulfill them.

Less than half of the documents (44.8\%) use the term “auditor” directly. For those who do, the role definition varies. Benbouzid et al. (\citeyear{benbouzid_pragmatic_2024}) and Chadda et al. (\citeyear{chadda_ai_2024}) describe how an auditor can be internal to an organization or an external third party in their protocol. In contrast, Wilson et al. (\citeyear{wilson_building_2021}) introduce the notion of a collaborative auditor, who is external to the audited organization but works with it to obtain information and access. Most often, the term is used without any indication of the auditor’s position, qualifications, or skill set. This ambiguity stems partly from the distributed nature of audit responsibilities. Nearly two-thirds of documents (65.5\%) describe \textit{joint auditing processes} in which multiple roles share oversight responsibilities, averaging three different auditor roles per document. Audits involving purposeful collaboration imply that gathering as many perspectives as possible promotes mutual learning, understanding, and communication \cite{liu_medical_2022, oala_ml4h_2020}. In a study auditing an AI-enabled radiology system, for instance, Mahajan et al. (\citeyear{mahajan_algorithmic_2020}) detail how performance must be examined by “a radiologist and a data scientist [who] need to sit together for hours at end and read every case where the AI failed…” to examine the implications of erroneous \ac{AI} (p. 133). As a result, who serves as “the auditor”, leads the process, and is responsible for the outcome in multi-party scenarios remains implicit. 

The choice of who is involved in an audit is also highly dependent on the level of access and knowledge. The \textit{developers} of the system being audited are involved in 46.6\% of audits, especially when access to the data and model components or technical intermediaries is required. This involvement is exemplified by Chadda et al.’s (\citeyear{chadda_ai_2024}) audit of named entity recognition technology, where data scientists and ML engineers must contribute to the evaluation process since they “train models and create the housing around them” (p. 23,037). Developers frequently audit inter-component sites.

In 44.8\% of documents, \textit{domain experts} are involved when expertise is required to make judgment calls, whether to formulate the initial audit expectations or interpret subjective results, especially when a contextual understanding of the system’s deployment is necessary. In Radiya-Dixit and Neff’s (\citeyear{radiya-dixit_sociotechnical_2023}) audit of \ac{FRT}, third-party experts are consulted to identify appropriate uses in practice. Surprisingly, in 39.7\% of documents, end-users assume lead auditing roles to report user-facing faults, applying co-audits \cite{gordon_co-audit_2023}, herd audits \cite{yang_game-theoretic_2023}, and end-user audits \cite{lam_end-user_2022}. Audits led by these actors primarily involve system-environment sites. 

\textit{Management roles} (involved in 10.3\% of audits) and policy-related roles (17.2\%) are prioritized in multi-system integration sites for organizational structuring, procurement, and \ac{AI} governance. Mökander et al. (\citeyear{mokander_blueprint_2025}) provide a clear example in their \textit{governance sub-audit} implicating management roles in data sourcing and documentation protocols for generative AI systems. We emphasize that this oversight extends beyond internal actors. 25.9\% of documents discuss situations wherein \textit{external vendors} (e.g., technology providers, manufacturers, and suppliers) are directly involved in audits, not to lead, but to assist with the audit trail. Ferrell and Anderegg (\citeyear {ferrell_applicability_2020}) rationalize external involvement as “true safety partner[s]” since “managing suppliers and integrating products from different sources are the most vulnerable points in safety management” (p. 6). Table~\ref{tab:summary} summarizes integration site characteristics.

\subsection{What the Audits Assess}

Having shown that system integration audits are emerging across a fragmented corpus, we now turn to \textit{what} they evaluate. In this section, fragmentation continues across the substance of these assessments: the qualities the audits claim to assess, the outputs they produce, and the functions those outputs serve. Rather than judging whether each audit achieves its stated objectives, we synthesize their purpose.

\subsubsection{Evaluated qualities}
The corpus shows substantial conceptual dispersion in what they aim to measure: we identify 33 distinct qualities, averaging seven per document, with one audit listing 18. Many documents use “umbrella terms” to nest multiple qualities. For instance, Helmer et al. (\citeyear{helmer_towards_2024}) use \textit{trust} to group six other qualities: \textit{fairness, autonomy, transparency, reliability, safety, and security}. The ten most popular qualities include: \textit{technical performance} (69.0\%), \textit{transparency} (56.9\%), \textit{accountability} (48.3\%), robustness (48.3\%), \textit{lack of bias} (44.8\%), \textit{consistency} (39.7\%), \textit{fairness} (31.0\%), \textit{process improvements} (29.3\%), \textit{trust} (27.6\%), and \textit{non-discrimination} (25.9\%). To a large extent, these results are consistent with the existing characterization of AI auditing as overly focused on technical performance or responsible AI principles \cite{koshiyama_towards_2024}. For instance, the subject of most performance-related evaluations is the AI \textit{model}, with measures such as \textit{accuracy} and \textit{specificity}. Yet, many documents equate these measures to overall \textit{system} performance. An example is DeGrave et al. (\citeyear{degrave_auditing_2025}), who claim to audit melanoma classification AI \textit{systems} but measure only \textit{model-level} metrics such as sensitivity, conflating model and system-level performance. Despite the dominance of popular audit qualities, three qualities specific to integration have been found.

\textbf{Compatibility (10.3\%)} evaluates whether intermediaries, the mechanisms connecting components and systems, operate in alignment. This typically involves technical checks at inter-component and multi-system sites, such as verifying synchronous software versions, as demonstrated by Helmer et al. (\citeyear{helmer_towards_2024}) in their audit catalogue for MLOps processes, or confirming that legacy systems remain operational for data transfer \cite{benbouzid_pragmatic_2024}. API interoperability is especially relevant. Sghaier et al (\citeyear{sghaier_verifiable_2024}), for instance, ensure their audit module’s API “conforms to the design principles of the REST architecture (REST API), facilitating user access to available AI models” (p. 2). Compatibility also extends to sociotechnical fit at system-environment sites, exemplified by Becker and Waltl (\citeyear{becker_auditing_2022}) who explicitly align audit activities with a company’s “digitization and data strategy”, “IT infrastructure goals”, and “cloud data hubs, lakes and warehouses” (p. 288).

\textbf{Completeness (8.6\%)} refers to whether the parts of a process contain the subparts needed to function properly. At inter-component sites, this could assess whether data entering a model are fully linked to all relevant categories required for prediction. For example, a heart disease risk model using country of origin would be complete if patient data could be associated with all relevant countries, rather than a limited subset. Knoblauch and Großmann (\citeyear{knoblauch_towards_2023}) describe this in technical terms, drawing on \ac{ISO} 25012 \textit{Data Quality Model's} use of “completeness” \cite{noauthor_international_2025-1}. At system-environment sites, completeness instead refers to whether the audit captures the range of deployment contexts in which the system operates. Hauer et al. (\citeyear{hauer_assuring_2021}), for instance, assess whether assurance cases, a common auditing approach, can account for relevant counterfactual conditions. Barker et al. (\citeyear{barker_feedbacklogs_2023}) similarly identify completeness as an audit objective: “Completeness: FeedbackLogs should provide comprehensive details about stakeholder feedback and subsequent practitioner updates” (p. 2). Notably, completeness is absent in multi-system sites, likely due to access constraints.

\textbf{Oversight (6.9\%)} describes the extent to which decision-making authority is clearly established and exercised across integration sites (Lam et al. 2022). It reflects how well supervisory roles are defined and empowered within an audit process, including the authority to uphold subjective audit judgments \cite{moreau_failing_2024}, govern responsibility over architectural elements \cite{boltz_human_2024}, sanction deployment decisions \cite{liu_medical_2022}, and determine remediation actions \cite{sloane_silicon_2022}. At inter-component sites, oversight concerns authority within the development pipeline. Polzer et al. (\citeyear{polzer_validation_2022}), for instance, audit how an explainable AI user-interaction module applies decision criteria to neural network outputs, clarifying whether system-identified regions can support authorized judgments. At system-environment sites, oversight concerns how authority is retained within operation. This is illustrated by Moreau et al. (\citeyear{moreau_failing_2024}) who examine how social workers adjust or maintain their own risk assessment scores relative to those produced by a decision-support algorithm. These sites may also involve oversight across the communication of audit findings to affected stakeholders, as in Radiya-Dixit and Neff’s (\citeyear{radiya-dixit_sociotechnical_2023}) audit scorecard assessing the disclosure of FRT trials by three United Kingdom police forces. At multi-system sites, oversight instead concerns authority across organizational boundaries. Mökander and Floridi (\citeyear{floridi_operationalising_2023}) highlight this in an ethics-based audit of AstraZeneca’s AI development, where external partners supplied training data and models. The audit assessed how vendor dependence limited the company’s and auditors’ ability to audit. For a summary of all qualities, their proxies, and proportions, see Appendix D (Table D1). 

\subsubsection{Audit outputs and functions}
Variation across the corpus also extends to the audit outputs. We identified nine result types: quantitative metrics (53.4\%), descriptive statements (41.4\%), binary/discrete outcomes (34.5\%), audit summaries/reports (19.0\%), question sets (17.2\%), comparative analyses (5.2\%), audit tasks/requirements (6.9\%), standards/certifications (3.4\%), and miscellaneous (3.4\%). On average, each audit produces three types. Rather than treating these outputs as isolated artifacts, we examine what each output enabled the audit to do. Through this interpretation, the outputs could be grouped into four recurring functions. As descriptive statements (e.g., an audited system produced biased outputs or showed inconsistent performance across contexts) summarized evidence across all functions, the output was not directly mapped to any one (see Table~\ref{tab:audit-result-outputs}).

\textbf{(1) Risk exploration} takes stock of vulnerabilities that may arise when an AI system is integrated. In the corpus, this function is reflected in question sets and comparative analyses. Question sets appear in 17.2\% of documents and encourage auditees, most often developers and end-users, to consider potential system development or deployment risks. For example, Naja et al. (\citeyear{naja_using_2022}) “gathered competency questions relating to accountability of AI systems" to construct a set of eight audit topics and their associated risks (p. 74,385). Comparative analyses appear less frequently (5.2\%) and examine potential AI system failures against domain expert performance, as in Oakden-Rayner et al. (\citeyear{oakden-rayner_validation_2022}), who compare AI classification failure modes with radiologists. These outputs serve an exploratory function by prompting reflexivity: question sets prompt stakeholders to identify potential failure points, while comparative analyses reveal discrepancies between AI performance and human judgment. Together, they help auditors determine which integration sites warrant closer examination and for which faults, including inter-component relationships between data and model artifacts and system-environment relationships involving human decision-making processes (Naja et al. \citeyear{naja_using_2022}, pp. 74,386-74,387). \textit{Risk assessments} are the most common exploratory instrument, acting as an initial scoping and diagnostic mechanism. Benbouzid et al. (\citeyear{benbouzid_pragmatic_2024}) and Liu et al. (\citeyear{liu_medical_2022}), for instance, begin their audits with risk assessments that map which elements (e.g., outputs) should be targeted. Both studies encourage broad stakeholder input to capture diverse risk perspectives. Mökander et al. (\citeyear{mokander_blueprint_2025}) similarly use risk assessments at multi-system sites, posing open-ended questions to explore upstream inter-component risks and downstream system-environment risks. 

\begin{table}[t]
\centering
\setlength{\tabcolsep}{3pt}

{\small
\begin{tabular}{|p{\dimexpr0.40\columnwidth-2\tabcolsep\relax}|p{\dimexpr0.60\columnwidth-2\tabcolsep-3\arrayrulewidth\relax}|}
\hline
\textbf{Audit Outputs} & \textbf{Related Function} \\
\hline

Question sets and comparative analyses &
Risk exploration--“Guiding” \\
\hline

Quantitative and binary/discrete outputs &
Risk determination--“Concretizing” \\
\hline

Audit tasks/requirements &
Coordination--“Organizing” \\
\hline

Audit summaries/reports and standards/certifications &
Procedural regularity--“Stabilizing” \\
\hline
\end{tabular}
}

\caption{Mapping audit outputs to associated functions. Note that the “descriptive statement” output covers all four functions due to its breadth and was thereby excluded.}
\label{tab:audit-result-outputs}
\end{table}

{\textbf{(2) Risk determination} describes the process of confirming or characterizing risks to establish their prevalence and severity. In this function, audits move beyond vulnerability identification to evaluate whether suspected risks manifest and warrant mitigation, restrictions, or redesign. Unlike risk exploration, which can involve a wide range of stakeholders, risk determination typically relies on developers or domain experts who have the technical access or contextual expertise necessary to establish appropriate evaluation thresholds \cite{gerchick_devil_2023}. In the corpus, two outputs serve a deterministic role. Quantitative outputs, such as performance measures and statistical tests, appear in 53.4\% of documents and assess whether predefined expectations are met. Binary/discrete outcomes, represented by pass/fail or categorical acceptance decisions, appear in 34.5\%. At inter-component sites, audits often employ \textit{traditional risk models}, since failures can be traced through sequential relationships among components (e.g., data-to-model, model-to-output). Adler and Klaes (\citeyear{adler_assurance_2022}), for instance, implement “different error propagation models like component fault trees” to govern cascading effects (p. 291). At system-environment sites, where explicit performance standards are less agreed upon, domain experts are relied upon to interpret findings and make evaluative audit judgments. In Gerchick et al. (\citeyear{gerchick_devil_2023}), workers reviewing a child welfare screening tool can accept or override the system’s risk scores, effectively deciding whether the model’s outputs should be accepted in practice. Risk determination also appears in multi-system contexts, although assessments are often separated into upstream and downstream evaluations rather than conducted directly across system boundaries. While \ac{GPAI} providers can evaluate risks associated with their own models, system integrators often lack upstream visibility and assess risks indirectly through their deployments.

\textbf{(3) Coordination} refers to the organization, distribution, and alignment of auditing activities across integration sites and their actors. Rather than identifying or confirming risks, this function establishes the operational workflows and communication channels through which information, responsibilities, and evidence are exchanged during the audit. In the corpus, coordination appears in audit task lists and requirement suites (6.9\%), which specify the roles and resources necessary for audit completion. As coordination concerns how audit work is organized, all three integration sites are implicated. Knoblauch and Großmann (\citeyear{knoblauch_towards_2023}), for example, coordinate simultaneous evaluation processes through technical intermediaries linking MLOps monitoring, audit APIs, and certification workflows, reflecting inter-component integration. Krafft et al. (\citeyear{krafft_action-oriented_2021}) instead develop process flow diagrams mapping interactions among system stakeholders, reflecting system-environment integration. Another dimension of coordination is building consensus around what falls outside the system boundary. As Wilson et al. (\citeyear{wilson_building_2021}) note, “just as important as defining what we were auditing is understanding what we were not auditing.” (p. 5) In their audit of the hiring platform Pymetrics, the authors restrict their evaluation to a single product, excluding coordinated efforts across additional tools due to access blockers.

\textbf{(4) Procedural regularity} formalizes consistent and traceable procedures for conducting and documenting audits. Whereas coordination aligns audit actors and activities, procedural regularity stabilizes the audit process itself by making evaluation steps and documentation practices repeatable over time. Exemplified by audit summaries/reports (19.0\%) and standards/certifications (3.4\%), these artifacts prescribe chains of reasoning across audit steps, decisions, and findings. They create reference points permitting the replication or incorporation of auditing practices into governance processes. Helmer et al. (\citeyear{helmer_towards_2024}), for instance, create audit decision records that “will guide the development of future software by prioritizing important decisions at the right level of the project maturity levels” (p. 6). Knoblauch and Großmann (\citeyear{knoblauch_towards_2023}) similarly produce standardized audit reports that regulatory bodies can use to assess compliance. A smaller subset emphasizes certification, usually as a hypothetical rather than operational mechanism, as in Becker and Waltl (\citeyear{becker_auditing_2022}) and Kazim et al. (\citeyear{kazim_systematizing_2021}), who outline how regulators \textit{could} audit an AI system end-to-end. 

\section{Discussion}

It is perhaps unsurprising that measuring the technical and responsible AI performance of a system is a core objective of the audits. However, we also discovered that integration-specific qualities, such as compatibility, completeness, and oversight, are reflected across the roles and four functions of an audit. Building on this, we now articulate three takeaways for institutionalizing system integration audits.

\subsection{Emerging Recognition of System Integration Risks in AI}

Motivated by the failures observed in the Boeing 737 MAX 8 incidents, this review began from the premise that a neglect of system integration evaluation can stem from both technical intermediaries and sociotechnical conditions. We therefore first located \textit{where} integration-related faults (see Table~\ref{tab:integration-faults}) can be situated in AI systems \cite{winter_system_2020}. Across the corpus, three central sites exemplify where these risks are recognized and, in some cases, addressed through audit design. At inter-component sites, the cascading nature of pipeline components feeding into each other, such as data-to-model relationships, can expose \textit{sequencing} faults. Adler and Klaes (\citeyear{adler_assurance_2022}) use error-propagation models, such as component-fault trees, to model these effects (Section 4.3). At system-environment sites, \textit{improper indication signalling} faults can be addressed where system use depends on transparent notification, interpretation, and institutional communication. Radiya-Dixit and Neff (\citeyear{radiya-dixit_sociotechnical_2023}), for example, audit disclosure mechanisms in police-enforced FRT trials (Section 4.3). Together, these audits demonstrate the potential to capture a range of integration-related risks. We emphasize that fault mitigation in multi-system sites was much harder to identify. 
Despite this review emphasizing GPAI risks and harms, \textit{less than 30\%} of the corpus audited these systems. For those that did, integration risks were separated into (a) upstream, technical model risks or (b) downstream, deployment risks. Rarely did providers and integrators work directly together to produce a complete trace of the AI supply chain between them. 

This work is nevertheless emerging. Hopkins et al. (\citeyear{hopkins_ai_2025}), for instance, quantitatively show how upstream model provider changes affect downstream task-specific systems by “[modelling] supply chains as directed graphs, where nodes correspond to AI components (and their corresponding actors)” (p. 1266). Such protocols could enable system integration audits to assess upstream and downstream integration risks \textit{simultaneously}. These examples, alongside the risk exploration and determination functions discussed, suggest a growing understanding of integration-related faults. Yet, their attempts in doing so remain largely implicit and fragmented in their approaches. Unlike aerospace regimes, system integration audits in AI have not yet matured to systematically address these faults. Few operationalizable qualities in the corpus specifically target integration-related risks, with compatibility, completeness, and oversight each appearing in less than 10\% of the corpus. This absence may reflect the continued dominance of component-level evaluation priorities on technical performance, even found within this corpus. These findings suggest a need for measurement processes targeting integration-related faults and qualities in AI. In many cases, measures may be adapted from existing practices. Lewis and Groth (\citeyear{lewis_metrics_2022}}), for example, identify 35 system-level software integration metrics, including \textit{ease of model notification} and \textit{framework coverage for known failure modes}, that could underpin specific audit criteria. Additional context-specific metrics may also be necessary. In time-sensitive healthcare services, for instance, \textit{server throughput capacity}--the total work a system can handle within a given period--could mitigate \textit{fail-safe margins} faults in clinical AI deployments \cite{assadi_integration_2022}.

\subsection{Comparison to Traditional Audit Expectations}
Given that our analysis centres on audits, which carry higher assurance expectations than AI evaluations, their legitimacy must be examined. In Section 2.1, we identified three audit conditions: independent auditors, systematic adherence to a standard or certification, and system-level evaluation of system-level risks. Given the review criteria, the third condition was readily met; the others were not.

First, most audits in the corpus lack an external, independent auditor. In established audit regimes, the certification, role, and expectations of external auditors are clearly defined. Under \ac{FAA} certification, for instance, third-party auditors must complete AS9100 Quality Management System training \cite{group_2026}. By contrast, over half of the documents in the corpus do not specify an explicit auditor role; likely reflecting the absence of enforceable standards in \ac{AI} auditing. More than ten distinct roles were involved in the audits, with more than two-thirds distributing joint evaluation responsibilities across multiple actors. These responsibilities vary by integration site based on expertise and access. While multi-perspective discourse has benefits, this arrangement weakens audit legitimacy. Most actors are neither external nor independent--the auditor effectively becomes the auditee \cite{falco_governing_2021}.

The second audit condition reveals similar shortcomings. While individual audits demonstrate procedural regularity, the corpus does not do so collectively. We identified more than 28 auditing approaches aimed at 33 distinct qualities. Despite many studies proposing requirements or reporting structures, fewer than 5\% reference an established standard or certification. Consequently, evaluation criteria vary significantly. For an emerging audit class, this goalpost shifting presents a major challenge. Provided that over 66\% of the audits are conceptualizations or proposals, these findings likely reflect the nascency of AI auditing research despite ongoing certification activities. But as they stand, system integration audits in AI more closely resemble evaluations than formal audits--which are not equivalent. In practice, many audits in the corpus leverage pre-existing AI evaluation tools, such as risk assessments, to gather evidence about component-level behaviours. We encourage the continued development of these audits to build on this rich landscape, serving as a coordinating mechanism capable of addressing gaps that existing tools cannot capture on their own.

\subsection{Information and Resource Access as a Constraint}
A recurring enabler or blocker in \textit{how} these audits are designed was their access to the audited system. As integration requires different components, systems, and intermediaries to be evaluated together, relevant elements may be distributed across different organizations, requiring mandatory coordination among the involved parties. Yet only a quarter of audits in the corpus discuss explicit vendor involvement. Nearly 75\% use an internal auditing approach, constrained by access to datasets, model parameters, or remedy avenues (Section 4.2). This may explain why developers, managers, and other internal actors play significant audit roles: they hold the technical or organizational access needed to conduct the audit. Even where access rights exist, as under the \ac{EU}'s \textit{General Data Protection Regulation}, weak enforcement makes such access difficult to obtain \cite{lin_more_2024}.

These access constraints do not hold in other audit domains. Ishibashi et al. (\citeyear{ishibashi_current_2024}), for example, show that the harmonization of the \textit{Medical Device Single Audit Program} enables economic agencies and regulatory authorities to have full access to audited systems. Mökander et al. (\citeyear{mokander_blueprint_2025}) elaborate that “this kind of privileged access is not unique. It is standard practice in governance audits in other fields. IT auditors, for example, have full access to material and reports related to operational processes and performance metrics” (p. 13). Intellectual property concerns exist across many industries, but they have not precluded audit access--the AI industry must not be an exception. These concerns will only escalate as large-scale GPAI models are integrated into diverse interfaces and environments, such as home robots. As in other safety-critical domains, downstream actors must have a full systems view of the integration faults, risks, and harms they may face. 

\subsection{Study limitations}
The term “system integration” is rarely used in the AI auditing community. To compensate for its absence, we used broader search criteria to keep important contributions, while narrowing the documents relevant to system integration. As queries were oriented to academic outlets and scientific databases, the screening process could under-represent industry, grassroots, and community work. Many documents also referenced Western-oriented laws, the OECD principles, and the EU AI Act, which constrain the representation of Global South deployments. 
%We remain mindful of these influences and reiterate the exploratory nature of this review, not confirmatory.

\section{Conclusion}
Over the last five years, system integration audits have emerged in AI, encompassing interactions among pipeline elements, external environments, and multi-system architectures. Yet these audits reveal ambiguity over who bears accountability and how assurance can be obtained. Given the corpus's limited empirical evidence, future work should use design ethnography to investigate how practitioners adopt, engage with, and resist these structures in practice. As AI systems become embedded in new applications, the field must prioritize dedicated approaches for integration-specific faults. As the Boeing 737 MAX 8 incidents made clear, system integration is not a peripheral concern; it can be a critical source of failure. It requires audits designed for the risks that arise when components, systems, and actors are brought together.

\bibliography{aaai2026}

\clearpage
\appendix

\renewcommand{\thetable}{A\arabic{table}}
\setcounter{table}{0}

\twocolumn[
\section{Completed Search Queries}
\label{app:queries}
\vspace{0.5 cm}

Table~\ref{tab:appendixA} reports the completed search conditions, queries, and number of document hits.

\vspace{0.75em}

\begin{center}
\normalsize
\setlength{\tabcolsep}{1mm}

\begin{tabular}{|p{0.15\textwidth}|p{0.22\textwidth}|p{0.47\textwidth}|p{0.10\textwidth}|}
\hline
\textbf{Database} & \textbf{Search Conditions} & \textbf{Exact Query} & \textbf{Hits} \\
\hline

\textbf{Articles from experts} &
Reached out to experts in their domain and personal inputs. &
Selected manually by domain experts and manually put into the corpus by reviewers. &
77 \\
\hline

\textbf{Scopus} &
Searched in the title and abstract fields. Only English. No retracted articles. &
(TITLE(((artificial W/1 intelligence) OR ``ai'' OR (machine W/1 learning) OR ``ml'' OR algorithmic OR algorithm) AND (audit OR auditing OR auditor)) OR ABS(((artificial W/1 intelligence) OR ``ai'' OR (machine W/1 learning) OR ``ml'' OR algorithmic OR algorithm) W/3 (audit OR auditing OR auditor))) AND (LIMIT-TO(LANGUAGE, ``English'')) AND (EXCLUDE(DOCTYPE, ``tb'')) &
1,086 \\
\hline

\textbf{EBSCOhost} &
Searched in the title and abstract fields. &
TI ((artificial w1 intelligence) or ``AI'' or (machine w1 learning) or ``ML'' or algorithmic or algorithm) AND TI (audit or auditing or auditor) OR AB ((artificial intelligence or ``ai'' or ``machine learning'' or ``ml'' or algorithmic or algorithm) w3 (audit or auditing or auditor)) &
615 \\
\hline

\textbf{Web of Science} &
Searched in the title and abstract fields. Applied the English language filter. &
Title(``artificial intelligence'' or ``ai'' or ``machine learning'' or ``ml'' or algorithmic or algorithm) AND Title(audit or auditing or auditor) OR Abstract(``artificial intelligence'' or ``ai'' or ``machine learning'' or ``ml'' or algorithmic or algorithm) NEAR/3 (audit or auditing or auditor) &
673 \\
\hline

\textbf{IEEE Xplore} &
Searched in the ``document title'' and ``abstract'' fields. &
``artificial intelligence'' OR ``ai'' OR ``machine learning'' OR ``ml'' OR algorithmic OR algorithm AND ``auditor'' OR audit OR auditing &
1,018 \\
\hline

\textbf{WorldCat} &
Only selected chapters for the data format. Applied the English language filter. No proximity operator possible. &
ti:((``artificial intelligence'' OR ``ai'' OR ``machine learning'' OR ``ml'' OR ``algorithmic'' OR ``algorithm'') AND (``audit'' OR ``auditing'' OR ``auditor'')) &
46 \\
\hline

\textbf{ProQuest} &
Searched in title without proximity term or abstract with the near proximity operator. Scholarly journals, working papers, trade journals, conference papers and proceedings, and books. &
title(((``artificial intelligence'' OR ``ai'' OR ``machine learning'' OR ``ml'' OR algorithmic OR algorithm) AND (audit OR auditing OR auditor))) OR abstract(((``artificial intelligence'' OR ``ai'' OR ``machine learning'' OR ``ml'' OR algorithmic OR algorithm) NEAR/3 (audit OR auditing OR auditor))) &
744 \\
\hline
\end{tabular}

\vspace{4pt}
\refstepcounter{table}
\label{tab:appendixA}
Table~\thetable: Final search terms, conditions, queries, and hits.
\end{center}

\vspace{1em}
]

\clearpage

\renewcommand{\thetable}{B\arabic{table}}
\setcounter{table}{0}

\twocolumn[
% \noindent{\Large\bf Appendix B: Documents Included in the Scoping Review}
% \par\vspace{0.5em}
\section{Documents Included in the Scoping Review}
\label{app:included articles}
\vspace{0.5 cm}
Tables B1 to B4 report the final list of documents included in the scoping review.

\vspace{0.75em}

\begin{center}
\normalsize
\setlength{\tabcolsep}{1mm}

\begin{tabular}{|p{0.10\textwidth}|p{0.84\textwidth}|}
\hline
\textbf{Document Number} & \textbf{Citation} \\
\hline
\textbf{1} & Adler, R., \& Klaes, M. (2022). Assurance Cases as Foundation Stone for Auditing AI-Enabled and Autonomous Systems: Workshop Results and Political Recommendations for Action from the ExamAI Project. In M. Rauterberg, F. Fui-Hoon Nah, K. Siau, H. Krömker, J. Wei, \& G. Salvendy (Eds.), HCI International 2022  -- Late Breaking Papers: HCI for Today’s Community and Economy (pp. 283 --300). Springer Nature Switzerland. \url{https://doi.org/10.1007/978-3-031-18158-0_21} \\
\hline
\textbf{2} & Kim, C., Gadgil, S. U., DeGrave, A. J., Cai, Z. R., Daneshjou, R., \& Lee, S.-I. (2023). Fostering transparent medical image AI via an image-text foundation model grounded in medical literature (p. 2023.06.07.23291119). medRxiv. \url{https://doi.org/10.1101/2023.06.07.23291119} \\
\hline
\textbf{3} & Alagarswamy, D. P., Berghoff, C., Danos, V., Langer, F., Markert, T., Schneider, G., Twickel, A. von, \& Woitschek, F. (2023). Towards Audit Requirements for AI-based Systems in Mobility Applications. Proceedings of the 9th International Conference on Information Systems Security and Privacy, 339 --348. \url{https://doi.org/10.5220/0011619500003405} \\
\hline
\textbf{4} & Helmer, L., Martens, C., Wegener, D., Akila, M., Becker, D., \& Abbas, S. (2024). Towards Trustworthy AI Engineering ---A Case Study on integrating an AI audit catalog into MLOps processes. Proceedings of the 2nd International Workshop on Responsible AI Engineering, 1 --7. \url{https://doi.org/10.1145/3643691.3648584} \\
\hline
\textbf{5} & Moreau, T., Sinatra, R., \& Sekara, V. (2024). Failing Our Youngest: On the Biases, Pitfalls, and Risks in a Decision Support Algorithm Used for Child Protection. The 2024 ACM Conference on Fairness, Accountability, and Transparency, 290 --300. \url{https://doi.org/10.1145/3630106.3658906} \\
\hline
\textbf{6} & Gursoy, F., \& Kakadiaris, I. A. (2022). System Cards for AI-Based Decision-Making for Public Policy (arXiv:2203.04754). arXiv. \url{https://doi.org/10.48550/arXiv.2203.04754} \\
\hline
\textbf{7} & Boltz, N., Getir Yaman, S., Inverardi, P., De Lemos, R., Van Landuyt, D., \& Zisman, A. (2024). Human empowerment in self-adaptive socio-technical systems. Proceedings of the 19th International Symposium on Software Engineering for Adaptive and Self-Managing Systems, 200 --206. \url{https://doi.org/10.1145/3643915.3644082} \\
\hline
\textbf{8} & Adel, S. M., Bichu, Y. M., Pandian, S. M., Sabouni, W., Shah, C., \& Vaiid, N. (2024). Clinical audit of an artificial intelligence (AI) empowered smile simulation system: A prospective clinical trial. Scientific Reports, 14(1), 19385. \url{https://doi.org/10.1038/s41598-024-69314-6} \\
\hline
\textbf{9} & Oakden-Rayner, L., Gale, W., Bonham, T. A., Lungren, M. P., Carneiro, G., Bradley, A. P., \& Palmer, L. J. (2022). Validation and algorithmic audit of a deep learning system for the detection of proximal femoral fractures in patients in the emergency department: A diagnostic accuracy study. The Lancet Digital Health, 4(5), e351 --e358. \url{https://doi.org/10.1016/S2589-7500(22)00004-8} \\
\hline
\textbf{10} & Liu, X., Glocker, B., McCradden, M. M., Ghassemi, M., Denniston, A. K., \& Oakden-Rayner, L. (2022). The medical algorithmic audit. The Lancet Digital Health, 4(5), e384 --e397. \url{https://doi.org/10.1016/S2589-7500(22)00003-6} \\
\hline
\textbf{11} & Mahajan, V., Venugopal, V. K., Murugavel, M., \& Mahajan, H. (2020). The Algorithmic Audit: Working with Vendors to Validate Radiology-AI Algorithms ---How We Do It. Academic Radiology, 27(1), 132 --135. \url{https://doi.org/10.1016/j.acra.2019.09.009} \\
\hline
\textbf{12} & Lam, M. S., Pandit, A., Kalicki, C. H., Gupta, R., Sahoo, P., \& Metaxa, D. (2023). Sociotechnical Audits: Broadening the Algorithm Auditing Lens to Investigate Targeted Advertising. Proceedings of the ACM on Human-Computer Interaction, 7(CSCW2), 1 --37. \url{https://doi.org/10.1145/3610209} \\
\hline
\textbf{13} & Dong, T., Li, S., Chen, G., Xue, M., \& Zhu, H. (2023). RAI2: Responsible Identity Audit Governing the Artificial Intelligence. Proceedings 2023 Network and Distributed System Security Symposium. Network and Distributed System Security Symposium. \url{https://doi.org/10.14722/ndss.2023.241012} \\
\hline
\textbf{14} & Oala, L., Fehr, J., Gilli, L., Balachandran, P., Leite, A. W., Calderon-Ramirez, S., Li, D. X., Nobis, G., Alvarado, E. A. M., Jaramillo-Gutierrez, G., Matek, C., Shroff, A., Kherif, F., Sanguinetti, B., \& Wiegand, T. (2020). ML4H Auditing: From Paper to Practice. Proceedings of the Machine Learning for Health NeurIPS Workshop, 280 --317. \url{https://proceedings.mlr.press/v136/oala20a.html} \\
\hline
\textbf{15} & Hauer, M. P., Adler, R., \& Zweig, K. (2021). Assuring Fairness of Algorithmic Decision Making. 2021 IEEE International Conference on Software Testing, Verification and Validation Workshops (ICSTW), 110 --113. \url{https://doi.org/10.1109/ICSTW52544.2021.00029} \\
\hline
\end{tabular}

\vspace{4pt}
\refstepcounter{table}
\label{tab:appendixB1}
Table~\thetable: Final list of documents included in the scoping review (Part 1).
\end{center}

\vspace{1em}
]

\begin{table*}[t]
\centering
\normalsize
\setlength{\tabcolsep}{1mm}
\begin{tabular}{|p{0.10\textwidth}|p{0.84\textwidth}|}
\hline
\textbf{Document Number} & \textbf{Citation} \\
\hline
\textbf{16} & Gordon, A. D., Negreanu, C., Cambronero, J., Chakravarthy, R., Drosos, I., Fang, H., Mitra, B., Richardson, H., Sarkar, A., Simmons, S., Williams, J., \& Zorn, B. (2023). Co-audit: Tools to help humans double-check AI-generated content (arXiv:2310.01297). arXiv. \url{https://doi.org/10.48550/arXiv.2310.01297} \\
\hline
\textbf{17} & Calacci, D., \& Pentland, A. (2022). Bargaining with the Black-Box: Designing and Deploying Worker-Centric Tools to Audit Algorithmic Management. Proceedings of the ACM on Human-Computer Interaction, 6(CSCW2), 1 --24. \url{https://doi.org/10.1145/3570601} \\
\hline
\textbf{18} & Rakova, B., \& Dobbe, R. (2023). Algorithms as Social-Ecological-Technological Systems: An Environmental Justice Lens on Algorithmic Audits. 2023 ACM Conference on Fairness, Accountability, and Transparency, 491 --491. \url{https://doi.org/10.1145/3593013.3594014} \\
\hline
\textbf{19} & Galdon Clavell, G., Martín Zamorano, M., Castillo, C., Smith, O., \& Matic, A. (2020). Auditing Algorithms: On Lessons Learned and the Risks of Data Minimization. Proceedings of the AAAI/ACM Conference on AI, Ethics, and Society, 265 --271. \url{https://doi.org/10.1145/3375627.3375852} \\
\hline
\textbf{20} & Naja, I., Markovic, M., Edwards, P., \& Cottrill, C. (2021). A Semantic Framework to Support AI System Accountability and Audit. In R. Verborgh, K. Hose, H. Paulheim, P.-A. Champin, M. Maleshkova, O. Corcho, P. Ristoski, \& M. Alam (Eds.), The Semantic Web (Vol. 12731, pp. 160 --176). Springer International Publishing. \url{https://doi.org/10.1007/978-3-030-77385-4_10} \\
\hline
\textbf{21} & Mökander, J., Curl, J., \& Kshirsagar, M. (2025). A blueprint for auditing generative AI. In Research Handbook on the Law of Artificial Intelligence (pp. 307-327). Edward Elgar Publishing. \url{https://arxiv.org/pdf/2407.05338?} \\
\hline
\textbf{22} & Lam, M. S., Gordon, M. L., Metaxa, D., Hancock, J. T., Landay, J. A., \& Bernstein, M. S. (2022). End-User Audits: A System Empowering Communities to Lead Large-Scale Investigations of Harmful Algorithmic Behavior. Proceedings of the ACM on Human-Computer Interaction, 6(CSCW2), 1 --34. \url{https://doi.org/10.1145/3555625} \\
\hline
\textbf{23} & Fernsel, L., Kalff, Y., \& Simbeck, K. (2024). Where Is the Evidence? A Plugin for Auditing Moodle’s Learning Analytics: Proceedings of the 16th International Conference on Computer Supported Education, 262 --269. \url{https://doi.org/10.5220/0012689800003693} \\
\hline
\textbf{24} & Dash, A., Chakraborty, A., Ghosh, S., Mukherjee, A., \& Gummadi, K. P. (2021). When the Umpire is also a Player: Bias in Private Label Product Recommendations on E-commerce Marketplaces. Proceedings of the 2021 ACM Conference on Fairness, Accountability, and Transparency, 873 --884. \url{https://doi.org/10.1145/3442188.3445944} \\
\hline
\textbf{25} & Sghaier, M. O., Hadzagic, M., \& Shahbazian, E. (2024). Verifiable Human Autonomy Teaming for NORAD C2 Operations. 2024 IEEE Conference on Cognitive and Computational Aspects of Situation Management (CogSIMA), 137 --144. \url{https://doi.org/10.1109/CogSIMA61085.2024.10553806} \\
\hline
\textbf{26} & Polzer, A., Fleiß, J., Ebner, T., Kainz, P., Koeth, C., \& Thalmann, S. (2022). Validation of AI-based Information Systems for Sensitive Use Cases: Using an XAI Approach in Pharmaceutical Engineering. Hawaii International Conference on System Sciences. \url{https://doi.org/10.24251/HICSS.2022.186} \\
\hline
\textbf{27} & Naja, I., Markovic, M., Edwards, P., Pang, W., Cottrill, C., \& Williams, R. (2022). Using Knowledge Graphs to Unlock Practical Collection, Integration, and Audit of AI Accountability Information. IEEE Access, 10, 74383 --74411. \url{https://doi.org/10.1109/ACCESS.2022.3188967} \\
\hline
\textbf{28} & Knoblauch, D., \& Großmann, J. (2023). Towards a Risk-Based Continuous Auditing-Based Certification for Machine Learning. The Review of Socionetwork Strategies, 17(2), 255 --273. \url{https://doi.org/10.1007/s12626-023-00148-w} \\
\hline
\textbf{29} & Gerchick, M., Jegede, T., Shah, T., Gutierrez, A., Beiers, S., Shemtov, N., Xu, K., Samant, A., \& Horowitz, A. (2023). The Devil is in the Details: Interrogating Values Embedded in the Allegheny Family Screening Tool. 2023 ACM Conference on Fairness Accountability and Transparency, 1292 --1310. \url{https://doi.org/10.1145/3593013.3594081} \\
\hline
\textbf{30} & Ma, W., Song, Y., Xue, M., Wen, S., \& Xiang, Y. (2024). The “Code” of Ethics: A Holistic Audit of AI Code Generators. IEEE Transactions on Dependable and Secure Computing, 21(5), 4997 --5013. \url{https://doi.org/10.1109/TDSC.2024.3367737} \\
\hline
\end{tabular}
\caption{Final list of documents included in the scoping review (Part 2).}
\label{tab:appendixB2}
\end{table*}

\begin{table*}[t]
\centering
\normalsize
\setlength{\tabcolsep}{1mm}
\begin{tabular}{|p{0.10\textwidth}|p{0.84\textwidth}|}
\hline
\textbf{Document Number} & \textbf{Citation} \\
\hline
\textbf{31} & Kazim, E., Koshiyama, A. S., Hilliard, A., \& Polle, R. (2021). Systematizing Audit in Algorithmic Recruitment. Journal of Intelligence, 9(3), Article 3. \url{https://doi.org/10.3390/jintelligence9030046} \\
\hline
\textbf{32} & Benbouzid, D., Plociennik, C., Lucaj, L., Maftei, M., Merget, I., Burchardt, A., Hauer, M. P., Naceri, A., \& Smagt, P. van der. (2024). Pragmatic auditing: A pilot-driven approach for auditing Machine Learning systems (arXiv:2405.13191). arXiv. \url{https://doi.org/10.48550/arXiv.2405.13191} \\
\hline
\textbf{33} & Mökander, J., \& Floridi, L. (2023). Operationalising AI governance through ethics-based auditing: An industry case study. AI and Ethics, 3(2), 451 --468. \url{https://doi.org/10.1007/s43681-022-00171-7} \\
\hline
\textbf{34} & Spyridou, P. (Lia), Djouvas, C., \& Milioni, D. (2022). Modeling and Validating a News Recommender Algorithm in a Mainstream Medium-Sized News Organization: An Experimental Approach. Future Internet, 14(10), Article 10. \url{https://doi.org/10.3390/fi14100284} \\
\hline
\textbf{35} & Barnard, P., Bautista, J. R., Krook, J., Liu, A., Menéndez, H., Schmidt, A., \& Sookoor, T. (2023). MACAIF: Machine Learning Auditing for Clinical AI Fairness. Proceedings of the First International Symposium on Trustworthy Autonomous Systems, 1 --4. \url{https://doi.org/10.1145/3597512.3597522} \\
\hline
\textbf{36} & Wang, S., Huang, S., Zhou, A., \& Metaxa, D. (2024). Lower Quantity, Higher Quality: Auditing News Content and User Perceptions on Twitter/X Algorithmic versus Chronological Timelines. Proceedings of the ACM on Human-Computer Interaction, 8(CSCW2), 1 --25. \url{https://doi.org/10.1145/3687046} \\
\hline
\textbf{37} & Kingsley, S., Sinha, P., Wang, C., Eslami, M., \& Hong, J. I. (2022). “Give Everybody [..] a Little Bit More Equity”: Content Creator Perspectives and Responses to the Algorithmic Demonetization of Content Associated with Disadvantaged Groups. Proceedings of the ACM on Human-Computer Interaction, 6(CSCW2), 1 --37. \url{https://doi.org/10.1145/3555149} \\
\hline
\textbf{38} & Barker, M., Kallina, E., Ashok, D., Collins, K., Casovan, A., Weller, A., Talwalkar, A., Chen, V., \& Bhatt, U. (2023). FeedbackLogs: Recording and Incorporating Stakeholder Feedback into Machine Learning Pipelines. Equity and Access in Algorithms, Mechanisms, and Optimization, 1 --15. \url{https://doi.org/10.1145/3617694.3623239} \\
\hline
\textbf{39} & Spielvogel, C. P., Haberl, D., Mascherbauer, K., Ning, J., Kluge, K., Traub-Weidinger, T., Davies, R. H., Pierce, I., Patel, K., Nakuz, T., Göllner, A., Amereller, D., Starace, M., Monaci, A., Weber, M., Li, X., Haug, A. R., Calabretta, R., Ma, X., \ldots{} Nitsche, C. (2024). Diagnosis and prognosis of abnormal cardiac scintigraphy uptake suggestive of cardiac amyloidosis using artificial intelligence: A retrospective, international, multicentre, cross-tracer development and validation study. The Lancet Digital Health, 6(4), e251 --e260. \url{https://doi.org/10.1016/S2589-7500(23)00265-0} \\
\hline
\textbf{40} & Benthall, S., \& Shekman, D. (2023). Designing Fiduciary Artificial Intelligence. Equity and Access in Algorithms, Mechanisms, and Optimization, 1 --15. \url{https://doi.org/10.1145/3617694.3623230} \\
\hline
\textbf{41} & Kaplan, L., \& Sapiezynski, P. (2024). Comprehensively Auditing the TikTok Mobile App. Companion Proceedings of the ACM Web Conference 2024, 1198 --1201. \url{https://doi.org/10.1145/3589335.3651260} \\
\hline
\textbf{42} & Koomthanam, A. J., Tripathy, A., Serebryakov, S., Nayak, G., Foltin, M., \& Bhattacharya, S. (2024). Common Metadata Framework: Integrated Framework for Trustworthy Artificial Intelligence Pipelines. IEEE Internet Computing, 28(3), 37 --44. \url{https://doi.org/10.1109/MIC.2024.3377170} \\
\hline
\textbf{43} & Wilson, C., Ghosh, A., Jiang, S., Mislove, A., Baker, L., Szary, J., Trindel, K., \& Polli, F. (2021). Building and Auditing Fair Algorithms: A Case Study in Candidate Screening. Proceedings of the 2021 ACM Conference on Fairness, Accountability, and Transparency, 666 --677. \url{https://doi.org/10.1145/3442188.3445928} \\
\hline
\textbf{44} & Groves, L., Metcalf, J., Kennedy, A., Vecchione, B., \& Strait, A. (2024). Auditing Work: Exploring the New York City algorithmic bias audit regime. The 2024 ACM Conference on Fairness, Accountability, and Transparency, 1107 --1120. \url{https://doi.org/10.1145/3630106.3658959} \\
\hline
\textbf{45} & Gaebler, J. D., Goel, S., Huq, A., \& Tambe, P. (2024). Auditing the Use of Language Models to Guide Hiring Decisions (arXiv:2404.03086). arXiv. \url{https://doi.org/10.48550/arXiv.2404.03086} \\
\hline
\end{tabular}
\caption{Final list of documents included in the scoping review (Part 3).}
\label{tab:appendixB3}
\end{table*}

\begin{table*}[t]
\centering
\normalsize
\setlength{\tabcolsep}{1mm}
\begin{tabular}{|p{0.10\textwidth}|p{0.84\textwidth}|}
\hline
\textbf{Document Number} & \textbf{Citation} \\
\hline
\textbf{46} & DeGrave, A. J., Cai, Z. R., Janizek, J. D., Daneshjou, R., \& Lee, S.-I. (2025). Auditing the inference processes of medical-image classifiers by leveraging generative AI and the expertise of physicians. Nature Biomedical Engineering, 9(3), 294 --306. \url{https://doi.org/10.1038/s41551-023-01160-9} \\
\hline
\textbf{47} & Bharadhwaj, H. (2022). Auditing Robot Learning for Safety and Compliance during Deployment. Proceedings of the 5th Conference on Robot Learning, 1801 --1806. \url{https://proceedings.mlr.press/v164/bharadhwaj22a.html} \\
\hline
\textbf{48} & Becker, N., \& Waltl, B. (2022). Auditing and Testing AI  -- A Holistic Framework. In V. G. Duffy (Ed.), Digital Human Modeling and Applications in Health, Safety, Ergonomics and Risk Management. Health, Operations Management, and Design (pp. 283 --292). Springer International Publishing. \url{https://doi.org/10.1007/978-3-031-06018-2_20} \\
\hline
\textbf{49} & Ovalle, A., Dev, S., Zhao, J., Sarrafzadeh, M., \& Chang, K.-W. (2022). Auditing Algorithmic Fairness in Machine Learning for Health with Severity-Based LOGAN (arXiv:2211.08742). arXiv. \url{https://doi.org/10.48550/arXiv.2211.08742} \\
\hline
\textbf{50} & Ferrell, U. D., \& Anderegg, A. H. A. (2020). Applicability of UL 4600 to Unmanned Aircraft Systems (UAS) and Urban Air Mobility (UAM). 2020 AIAA/IEEE 39th Digital Avionics Systems Conference (DASC), 1 --7. \url{https://doi.org/10.1109/DASC50938.2020.9256608} \\
\hline
\textbf{51} & Rhea, A. K., Markey, K., D’Arinzo, L., Schellmann, H., Sloane, M., Squires, P., Arif Khan, F., \& Stoyanovich, J. (2022). An external stability audit framework to test the validity of personality prediction in AI hiring. Data Mining and Knowledge Discovery, 36(6), 2153 --2193. \url{https://doi.org/10.1007/s10618-022-00861-0} \\
\hline
\textbf{52} & Krafft, P. M., Young, M., Katell, M., Lee, J. E., Narayan, S., Epstein, M., Dailey, D., Herman, B., Tam, A., Guetler, V., Bintz, C., Raz, D., Jobe, P. O., Putz, F., Robick, B., \& Barghouti, B. (2021). An Action-Oriented AI Policy Toolkit for Technology Audits by Community Advocates and Activists. Proceedings of the 2021 ACM Conference on Fairness, Accountability, and Transparency, 772 --781. \url{https://doi.org/10.1145/3442188.3445938} \\
\hline
\textbf{53} & Chadda, A., McGregor, S., Hostetler, J., \& Brennen, A. (2024). AI Evaluation Authorities: A Case Study Mapping Model Audits to Persistent Standards. Proceedings of the AAAI Conference on Artificial Intelligence, 38(21), Article 21. \url{https://doi.org/10.1609/aaai.v38i21.30346} \\
\hline
\textbf{54} & Ema, A., Sato, R., Hase, T., Nakano, M., Kamimura, S., \& Kitamura, H. (2023). Advancing AI Audits for Enhanced AI Governance (arXiv:2312.00044). arXiv. \url{https://doi.org/10.48550/arXiv.2312.00044} \\
\hline
\textbf{55} & Radiya-Dixit, E., \& Neff, G. (2023). A Sociotechnical Audit: Assessing Police Use of Facial Recognition. 2023 ACM Conference on Fairness Accountability and Transparency, 1334 --1346. \url{https://doi.org/10.1145/3593013.3594084} \\
\hline
\textbf{56} & Sloane, M., Moss, E., \& Chowdhury, R. (2022). A Silicon Valley love triangle: Hiring algorithms, pseudo-science, and the quest for auditability. Patterns, 3(2). \url{https://doi.org/10.1016/j.patter.2021.100425} \\
\hline
\textbf{57} & Pettet, G., West, J., Robert, D., Khetani, A., Kumar, S., Golla, S., \& Lavis, R. (2023). A retrospective audit of an artificial intelligence software for the detection of intracranial haemorrhage used by a teleradiology company in the United Kingdom. BJR|Open, 6(1), tzae033. \url{https://doi.org/10.1093/bjro/tzae033} \\
\hline
\textbf{58} & Yang, Y.-T., Zhang, T., \& Zhu, Q. (2023). A Game-Theoretic Analysis of Auditing Differentially Private Algorithms with Epistemically Disparate Herd. In J. Fu, T. Kroupa, \& Y. Hayel (Eds.), Decision and Game Theory for Security (pp. 349 --368). Springer Nature Switzerland. \url{https://doi.org/10.1007/978-3-031-50670-3_18} \\
\hline
\end{tabular}
\caption{Final list of documents included in the scoping review (Part 4).}
\label{tab:appendixB4}
\end{table*}

\clearpage
\section{Description of AI System Elements}
\label{app:descriptions}
\vspace{0.5 cm}
We initially identified four core stages of an AI system: data/input, model, output, and user-interaction. These stages, as defined by the OECD’s definition of an AI system, were deductively included in the analysis.

\textbf{Data/input:} Among these stages, 87.9\% of articles explicitly included the data/input stage, including datasets, data sources, data sheets, data dictionaries, training, testing, and validation data, metadata or provenance data, synthetic data use, data structures, and input prompts and parameters. This stage also comprises characteristics of the data, including its format, modalities, properties, scale, and availability and representation. Other areas, such as data sharing agreements, data storage, and privacy, were also covered, often related to data protection laws. This stage often overlapped with others, including output data, production data, deployment data, and even audit data. The actions surrounding the data stage were wide, including data collection and curation, annotation and labelling, pre-processing, importing and uploading, maintenance, transfer and transmission, identification or de-identification, sharing, versioning and updating, infringement, extraction, and general handling. Many aspects of prompt engineering also fell into this stage.

\textbf{Model:} 67.2\% of articles explicitly discussed the model stage. The codes in this stage ranged from discussions surrounding model weights, parameters, features, specifications, decision thresholds, proxies, inferences, and both loss and objective functions. A variety of model types and algorithm choices were reported, along with more generalized designs, configurations, decision structures and logic, architectures, techniques, and trade-offs. Model confidence, behaviour, constraints, capabilities, limitations, internal reasoning, inference, realization, calibration, and alignment were also included, in addition to model creation (manual or automated), selection, pre- and post-training, testing, under- and over-fitting, fine-tuning, comparison, transmission, reuse, extraction, licensing (e.g., copyright and intellectual property), updating, versioning, and optimization processes. Artifacts such as model cards were also tagged, along with intermediaries such as model plug-ins, microservices, API licensing agreements, and the ``housing'' surrounding models.

\textbf{Output:} Within the output stage, which 65.5\% of articles explicitly discussed, various result types and formats, findings, and evidence were discussed to ultimately support further decisions. This translation took many forms, including metrics, measures, thresholds, targets, standards, aims, scores, performance levels, intervention points, criteria, counterfactuals, output distributions, and more broadly, outcomes, confirmations, and conclusions. This stage also predominantly captured feedback, including predictions, probabilities, underlying mechanisms, conditions, observations, responses, error checks, implications, and the use of the output to full screen. Comparative terms were often used to describe the relative positioning of an output, such as similarities, differences, adherence, changes in response, causation, correlations, aggregations, ``net effects,'' or the extent or impact of the measurement process. The term ``claim validation'' was also used. Various actions took place within the output, such as detection, revealing, interpretation, communication, quantification or discretization, demonstrations, exhibitions, prognosis, yields, and ultimately, summarizations resulting in some form of a product. Various tools were used to display outputs, including dashboards and visualizations.

\textbf{User-interaction:} 48.3\% of articles referenced the user-interaction stage. This stage concerns user involvement, whether actively applied in real time or passively implicated. It detailed user perceptions, behaviours, beliefs, reactions, sentiments, heuristics, preferences, desires, expectations, profiles, and characterizations relating to the audit, as well as usage and knowledge workflows. Both individual and collective user action were discussed. For instance, several audits leveraged user-interaction through user-led audits or co-audits, where users conducted an audit, or generated data for the audit, by themself or with assistance from an AI system. Human decision-making and human-AI synergy were also prominent collaboration themes, including human-out-of-the-loop or human-in-command; a note on anthropomorphism was also recorded. Conversations about information-gathering and retrieval processes, including human feedback and reporting, and communication; approval and delivery channels, whether direct and purposeful or indirect; and how user experience can benefit from various acclimation, customization, personalization, management, and optimization approaches, such as intent matching, were also common. Most of this discussion surrounded user disclosure, awareness, exposure, and insight. Some of this was enacted through user enrollment, onboarding, training, and general guidance and instruction. It also touched on differences among user interfaces, including web, mobile, and physical applications (e.g., embodied and robotic cases), and how information was presented and subsequently ingested by users.

Intermediaries that enabled or permitted users to take specific actions, such as API access, wrappers, microservices, emulators or simulators, various output templates, dashboards, visuals, and displays, as well as user agreements and contracts, were also addressed. User data, such as network traffic surveillance, surveys, comments, requests, and history, were also discussed. Action words such as interpretation, expression, and routine computing activities, such as browsing, toggling, updating, uploading, or downloading, were noted. Unsurprisingly, there was overlap between the output stage, given that most user-interaction centres on the AI system’s output. Clear examples of this included user-centred measures.

\textbf{Deployment:} The deployment stage, discussed in 51.7\% of articles, refers to the continuous, real-time, and active implementation, operation, monitoring, and maintenance of an AI system, once it is installed, configured, or released into a user-facing environment. Often, this stage was referred to as the final verification step. Contextual fit, appropriateness, and transferability were repeatedly emphasized to assess whether a system’s behaviour performs as intended across different times, settings, and user-interactions. Hauer et al. describe deployment as ``the final, social process in which the decision making is embedded needs to be checked'' \cite{hauer_assuring_2021}. Distribution shifts arising from evolving data, models, and user interactions were central, followed by discussions of acceptable evaluation deviation under real-world conditions. As a result, articles primarily discussed the deployment stage in a retrospective sense, detailing remedy, recourse, restoration, and rollback functions if deviations became too significant. Dependencies on external services, such as cloud computing, were also mentioned to ensure it can be deployed to production. Deployment methods included sandboxing, post-market surveillance and analysis, scenario testing, ethnographic techniques, and feasibility studies.

\textbf{Ecosystem:} We additionally note a broader stage, which we labelled as the ecosystem. Twelve articles directly used the term to position themselves in favour of a more human- and environment-centric ``ecosystem view'' such as Groves et al. (2024, p. 1115). We follow Barker et al.’s definition, who describe the ecosystem as ``the socio-technical realm in which the ML pipeline lives'' (2023, p. 5). This stage, or perhaps realm, was dominant, reflected in 77.6\% of articles. The scope being discussed ranged from AI system ecosystems to auditing ecosystems, as well as more domain-specific ecosystems (e.g., news or hiring ecosystems). This stage was more systemic, often focusing on longer-term risks, harms, and considerations related to the bigger picture of an AI system’s viability, sustainability, and longevity. This discussion included compounding and summative impacts to produce a road map for subsequent activities. Oftentimes, this led to discussions of organizational, management, and governance structures surrounding the AI system, as well as of how standards and laws could promote or impede procedural regularity. Power, control, and access dynamics were frequently mentioned, including sub-themes of relational dynamics; information and knowledge flows; and, more generally, responsible AI culture. These dynamics also included discussions of information, knowledge, and material flows, especially surrounding supply chains and their actors. Factors enabling or blocking AI system audits were therefore identified in this stage, including socioeconomic status; cost, incentive, and market structures; political priorities and climates; lobbying and crowd-sourcing efforts; and historical and cultural positioning. The question of how the AI system was positioned (e.g., at whose benefit and expense) was central to the ecosystem discussion, including the assumptions, motives, intentions, and rationales of the various roles involved. Folk theories were used to explicate normative value decisions and further sense-making, acting as the foundation for many ecosystem spheres.

\clearpage

\renewcommand{\thetable}{D\arabic{table}}
\setcounter{table}{0}

\twocolumn[
\section{Evaluated Qualities}
\label{app:qualities}
\par\vspace{0.5em}

Table~\ref{tab:appendixD} reports the assessed qualities, corresponding proxies, and proportions of each present in the corpus.

\vspace{0.75em}

\begin{center}
\normalsize
\setlength{\tabcolsep}{1mm}

\begin{tabular}{|p{0.22\textwidth}|p{0.52\textwidth}|p{0.18\textwidth}|}
\hline
\textbf{Main Quality} & \textbf{Proxies}& \textbf{Percentage of Documents in the Corpus}\\
\hline
\textbf{Technical performance} & Accuracy and capability & 69.0\% \\
\hline
\textbf{Transparency} & Traceability, visibility, and provenance & 56.9\% \\
\hline
\textbf{Accountability} & Assurance & 48.3\% \\
\hline
\textbf{Robustness} & N/A & 48.3\% \\
\hline
\textbf{Lack of bias} & N/A & 44.8\% \\
\hline
\textbf{Consistency} & Reproducibility, reliability, stability, repeatability, and justifiability & 39.7\% \\
\hline
\textbf{Fairness} & N/A & 31.0\% \\
\hline
\textbf{Process improvements} & Scalability, improving the speed, reducing the resources, streamlining the communication of the evaluation, and improving efficiency & 29.3\% \\
\hline
\textbf{Trust} & Reputability and credibility & 27.6\% \\
\hline
\textbf{Non-discrimination} & Promoting diversity, equity, inclusivity, and representivity & 25.9\% \\
\hline
\textbf{Validity} & Verifiability and correctness & 25.9\% \\
\hline
\textbf{Safety} & N/A & 22.4\% \\
\hline
\textbf{Security} & N/A & 20.7\% \\
\hline
\textbf{Privacy} & N/A & 19.0\% \\
\hline
\textbf{Explainability} & Interpretability & 19.0\% \\
\hline
\textbf{Usefulness} & Efficacy, effectiveness from a use perspective, functionality, voracity, satisfactory, and reducing user burden & 17.2\% \\
\hline
\textbf{Harm recourse and redress} & N/A & 15.5\% \\
\hline
\textbf{Auditability} & Inspectability, examinability, reviewability, factors of effort and complexity in the auditing process, and feasibility & 12.1\% \\
\hline
\textbf{Meaningfulness} & Contextual fit and appropriateness & 10.3\% \\
\hline
\textbf{Compatibility} & Agreement, solidarity, and alignment & 10.3\% \\
\hline
\textbf{Justice} & Agreement, solidarity, and alignment & 8.6\% \\
\hline
\textbf{Completeness} & Extensiveness & 8.6\% \\
\hline
\textbf{Accessibility} & Readability, legibility, and literacy to non-technical users & 8.6\% \\
\hline
\textbf{Autonomy} & N/A & 6.9\% \\
\hline
\textbf{Ownership} & Agency & 6.9\% \\
\hline
\textbf{Oversight} & Awareness & 6.9\% \\
\hline
\textbf{Usability} & User appeal and aspects of personalization & 6.9\% \\
\hline
\textbf{Lawfulness} & Legality, care, prudence, and due diligence in a legal context & 6.9\% \\
\hline
\textbf{Adaptability} & Flexibility, ascertainability, and practicality in an auditing system & 3.4\% \\
\hline
\textbf{Currentness} & Recency & 3.4\% \\
\hline
\textbf{Sustainability} & N/A & 1.7\% \\
\hline
\textbf{Beneficence} & N/A & 1.7\% \\
\hline
\textbf{Non-maleficence} & N/A & 1.7\% \\
\hline
\textbf{Unspecified} & N/A & 1.7\% \\
\hline
\end{tabular}

\vspace{4pt}
\refstepcounter{table}
\label{tab:appendixD}
Table~\thetable: Assessed qualities, selected proxies, and corresponding proportions.
\end{center}

\vspace{1em}
]

\end{document}